\documentclass[prb,twocolumn]{revtex4}
\usepackage{amsmath}
\usepackage{graphicx}
\usepackage{amssymb}
\usepackage{amsthm, amscd}
\usepackage{amsfonts}
\usepackage{cancel}
\usepackage{times}
\usepackage{pstricks}
\usepackage{wasysym}
\usepackage{ulem}
\usepackage[dvips]{epsfig}
\usepackage{subfigure}
\usepackage{feynmf}

\newcommand{\dslash}{\not{\hbox{\kern-2pt $\partial$}}}
\newcommand{\bq}{\begin{equation}} 
\newcommand{\eq}{\end{equation}}
\newcommand{\bqa}{\begin{eqnarray}} 
\newcommand{\eqa}{\end{eqnarray}}
\newcommand{\nn}{\nonumber \\}
\newcommand{\bw}{\begin{widetext}}
\newcommand{\ew}{\end{widetext}}
\newcommand{\EGZ}{$\eta \rightarrow 0$~}
\begin{document}


\title{Low energy effective theory of Fermi surface coupled with U(1) gauge field  in 2+1 dimensions}

\author{Sung-Sik Lee}
\affiliation{Department of Physics $\&$ Astronomy, 
McMaster University,
Hamilton, Ontario L8S 4M1, 
Canada}

\date{\today}

\begin{abstract}

We study the low energy effective theory 
for a non-Fermi liquid state in 2+1 dimensions, 
where a transverse U(1) gauge field 
is coupled with a patch of Fermi surface 
with $N$ flavors of fermion
in the large $N$ limit.
In the low energy limit, 
quantum corrections are classified
according to the genus of the 2d surface 
on which Feynman diagrams can be drawn without a crossing
in a double line representation, 
and all planar diagrams are important in the leading order.
The emerging theory has the similar structure to 
the four dimensional SU(N) gauge theory in the large $N$ limit.
Because of strong quantum fluctuations 
caused by the abundant low energy excitations near the Fermi surface,
low energy fermions  remain strongly coupled 
even in the large $N$ limit.
As a result, there are infinitely many quantum corrections
that contribute to the leading frequency dependence of 
the Green's function of fermion on the Fermi surface.
On the contrary, the boson self energy is not modified 
beyond the one-loop level 
and the theory is stable in the large $N$ limit.
The non-perturbative nature of the theory also 
shows up in correlation functions
of gauge invariant operators.

\end{abstract}

\maketitle

\section{Introduction.}

Understanding non-Fermi liquid states 
is one of the central problems in condensed matter physics.
One way of obtaining non-Fermi liquids 
is to couple a Fermi surface with
a massless boson.
If a massless boson is 
associated with criticality 
achieved by fine tuning of microscopic parameters,
a non-Fermi liquid state arises 
at a quantum critical point.
Such non-Fermi liquid states have been observed
in heavy fermion compounds
near magnetic quantum critical points\cite{LOHNEYSEN,COLEMAN,GEGENWART}.
On the other hand, a boson can be dynamically tuned to
be massless, and a non-Fermi liquid state may occur 
within a finite parameter space.
The latter case may arise in 
the half filled Landau level of quantum Hall systems\cite{HALPERIN}
and some spin liquid states\cite{PALEE}.

Non-Fermi liquid states in 2+1 dimensions are of particular interest.
Experimentally, high temperature superconductors which exhibit
non-Fermi liquid behaviors in the strange metallic normal state\cite{PALEE}
are quasi two-dimensional.
There also exist two dimensional frustrated magnets\cite{SHIMIZU1,ITOU} 
whose ground states may be related to
a non-Fermi liquid state of 
fermionic spinons 
which carry only spin half but no charge\cite{ANDERSON}.
In the spin liquid state, fermionic spinons form a Fermi surface 
which is minimally coupled with an emergent U(1) gauge field\cite{MOTRUNICH,Lee_U1}.
The transverse component of the gauge field remains gapless at low energies
because it is not fully screened by particle-hole excitations.
The long-range interaction mediated by the 
transverse gauge field leads to a non-Fermi liquid state\cite{HALPERIN,PLEE92,POLCHINSKY,KIM94,NAYAK,ALTSHULER}.
The same low energy effective theory can arise in various 
microscopic models, such as 
frustrated boson systems\cite{MotrunichFisher}.
Fermi surface of spinless charged fermion
coupled with U(1) gauge field has been also proposed
for underdoped cuprates\cite{HolonMetal}.
On the theoretical side, 
two space dimension is special in that 
it is high enough to have an extended Fermi surface,
while it is low enough to support strong
quantum fluctuations in the low energy limit.
As a result of strong quantum fluctuations 
and infinitely many gapless excitations on an extended Fermi surface,
it is expected that a non-trivial interacting quantum field theory,
which is very different from relativistic quantum field theories,
can emerge in the low energy limit.
Even if the non-Fermi liquid state turns out to be unstable
against other more conventional states\cite{Lee_amp,Galitski},
the physics within a significant temperature range will
be inherited from the unstable non-Fermi liquid state, 
and it is still important to understand the parent non-Fermi liquid state.

Considerable studies have been devoted to the 
low energy effective theory of Fermi surface 
coupled with U(1) gauge field in 2+1D\cite{POLCHINSKY,KIM94,NAYAK,ALTSHULER,MotrunichFisher}.
In this system,
there is no controllable parameter
other than the number of fermion flavors $N$.
Therefore it is natural to 
attempt to develop a perturbative expansion 
in terms of $1/N$ ($N$) in the large (small) $N$ limit\cite{POLCHINSKY,KIM94,ALTSHULER,MotrunichFisher}. 
Based on the computation of some leading order diagrams,
it has been suggested that the $1/N$ expansion is well defined
and the low energy limit is described by 
a stable interacting theory in the large $N$ limit.
The purpose of this paper is 
to study the low energy effective theory
of the non-Fermi liquid state more systematically
in the large $N$ limit.
The key result of the paper is that
the theory is not in a perturbative regime
even in the large $N$ limit
because there are infinitely 
many leading order quantum corrections
for vertex functions of fermions 
residing on the Fermi surface.
This conclusion has been reached by 
a systematic classification 
of quantum corrections in the $1/N$ expansion.
Strong quantum fluctuations associated with 
the infinitely many gapless excitations
and the absence of the Lorentz symmetry
make the classification very different from 
relativistic quantum field theories.
The theory remains strongly coupled in the low energy limit,
and even the leading order quantum corrections can not be summed 
in a closed Dyson equation which can be truncated 
with a finite number of vertex corrections.

The paper is organized in the following way.
In Sec. II, we start by constructing 
a minimal local action (given in Eq. (\ref{a3}))
that captures the universal low energy physics of the
2+1 dimensional non-Fermi liquid state.
The minimal action is a renormalizable theory through which
one can probe the universal low energy physics at any (finite) energy scale
by sending all UV cut-off and cross-over scales to infinity.
This makes the analysis of the low energy physics more transparent, 
because all non-universal elements of the theory have been stripped away
from the minimal action. 
The peculiar property of the present non-relativistic theory 
with Fermi surface is that all local time derivative
terms are irrelevant in the low energy limit.
Nonetheless one can not completely drop the time derivative terms from
the bare action because 
if one does so,
the theory will not have any dynamics.
Therefore one needs to consider a special low energy limit
to retain non-trivial dynamics 
while keeping only universal properties of the low energy theory.
After we discuss the low energy limit of the minimal theory,
in Sec. III we classify quantum corrections in the large $N$ limit.
In Sec. III (a), we show that 
a naive $1/N$ expansion does not work
because power of a Feynman graph in $N$
is enhanced in the low energy limit.
This is due to strong quantum fluctuations enhanced by
abundant gapless particle-hole excitations 
near the Fermi surface.
More specifically, 
the leading frequency dependence of the fermion propagator
is of the order of $1/N$.
Due to the suppressed frequency dependence, 
magnitudes of Feynman diagrams are
enhanced whenever there exists a channel 
for virtual particle-hole excitations
to remain on the Fermi surface.
Based on this observation, 
in Sec. III (b), we show that general Feynman diagrams 
are classified according to the genus of a 2d surface on which
Feynman diagrams are drawn without a crossing
in a double line representation.
Here the double line representation is useful, 
not because gauge boson or fermions 
carry a doubled quantum number, 
but because it allows one to easily count in how many ways
particle-hole excitations can remain on the Fermi surface.
In particular, when a fermion is on the Fermi surface,
the fermion propagator is enhanced to the order of $N$
due to the suppressed frequency dependence.
If there are $n$ closed single line loops 
in the double line representation, 
all virtual particle-hole excitations can remain right on the
Fermi surface no matter what $n$ components of
internal momenta are. 
The abundant low lying excitations 
give rise to an enhancement factor with a positive power of $N$
in proportion to the number of independent channels
via particle-hole excitations remain on the Fermi surface.
Due to the enhancement factor, 
there exist infinitely many leading order quantum corrections 
for vertex functions of fermions on the Fermi surface.
In particular, one has to sum over infinitely many planar diagrams
to compute the leading frequency dependence of the fermion propagator.
Although fermions on the Fermi surface are strongly coupled,
the boson propagator is not modified beyond the one-loop level
in the large $N$ limit due to a kinematical constraint.

The genus expansion of Feynman diagrams 
in the present non-Fermi liquid state 
is very similar to 
that of the four dimensional SU(N) gauge theory 
in the large $N$ limit\cite{tHOOFT}.
In both theories, strong quantum fluctuations
make all planar diagrams to 
contribute to the quantum effective action
in the leading order of the $1/N$ expansion.
However, the physical origins for 
strong quantum fluctuations
are very different between the two theories.
In the SU(N) gauge theory, it is due to 
fluctuations of color degrees of freedom in the internal space,
while in the present theory, it is due to
fluctuations of the extended Fermi surface in the momentum space.

In Sec. IV, we study the dynamical properties of the theory 
in the large $N$ limit.
It is shown that there is no UV divergence in individual planar diagrams.
As a result, the theory is stable, 
and there is no quantum correction to the scaling dimension of fermion 
beyond one-loop order
if the summation of individually finite planar diagrams are finite.
In Sec. V, we discuss how the non-perturbative nature of the theory manifests
itself in correlation functions of a gauge invariant operator.

\section{Minimal theory and Low energy limit}

\subsection{Minimal local action}

We consider $N$ flavors of fermion with Fermi surface
coupled with a U(1) gauge boson in 2+1D.
In the low energy limit, fermions 
whose velocities are not parallel 
or anti-parallel to each other 
are essentially decoupled because 
1) fermions are strongly coupled only with
the boson whose momentum is perpendicular
to the Fermi velocity for a kinematic constraint, and
2) the angle that parametrizes Fermi surface
acquires a positive
anomalous scaling dimension,
becoming a decompactified variable which runs
from $-\infty$ to $\infty$ in the low energy limit\cite{LEE2008}.
As a result, two fermions which have different Fermi velocities
can not interact with each other
through any finite number of scatterings 
with the boson
in the low energy limit\cite{FL}. 
Therefore, in the low energy effective theory, 
it is justified to focus fermionic excitations 
locally in the momentum space.
In general, one has to consider all patches 
in which Fermi velocities are parallel or anti-parallel
to each other because all of them are strongly 
coupled with the boson in the same momentum region.

\begin{figure}
        \includegraphics[height=4cm,width=5cm]{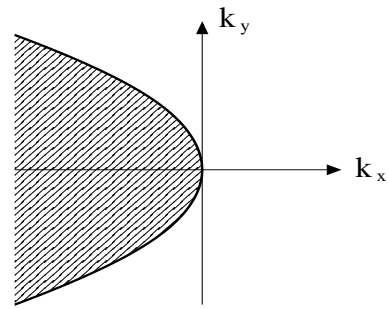} 
\caption{
The parabolic Fermi surface of the model in Eq. (\ref{a1}).
The shaded region includes negative energy states.
}
\label{fig:FS}
\end{figure}

In this paper, we will focus on low energy fermions
near one patch in the momentum space.
As we will see, understanding low energy dynamics 
in this simplified case
is already non-trivial.
At the end, we will comment on 
the applicability of this restricted theory,
and an extension to general cases which include 
other patches with opposite Fermi velocity.
We consider the Lagrangian density,
\bqa
{\cal L} & = & \sum_j 
\psi^*_{j} ( \partial_\tau - i v_x \partial_x - v_y \partial_y^2 ) \psi_{j} \nn
&& + \frac{e}{\sqrt{N}} \sum_j  a \psi^*_{j} \psi_{j} \nn
&& + a \left[  -\partial_\tau^2 - \partial_x^2 - \partial_y^2 \right] a,
\label{a1}
\eqa
where $\psi_{j}$ is the fermion of flavor $j=1,2,...,N$.
We have chosen the Fermi velocity to be along the $x$-direction at ${\bf k}=0$.
$v_x$ is the Fermi velocity and $v_y \sim \frac{1}{m}$ determines 
the curvature of the Fermi surface.
The Fermi surface is on $v_x k_x +  v_y k_y^2 = 0$ as is shown in Fig. \ref{fig:FS}.
This is a `chiral Fermi surface' where 
the x-component of Fermi velocity is always positive.
$a$ is the transverse component of an emergent U(1) gauge boson
in the Coulomb gauge $\nabla \cdot {\bf a} = 0$.
We ignore the temporal component of the gauge field which
is screened to a short range interaction.
The transverse gauge field is massless without a fine tuning
due to an emergent U(1) symmetry 
associated with the dynamical suppression of instantons\cite{IOFFE,LEE2008}.
$e$ is the coupling between fermions and the critical boson.

\begin{figure}
        \includegraphics[height=3cm,width=4cm]{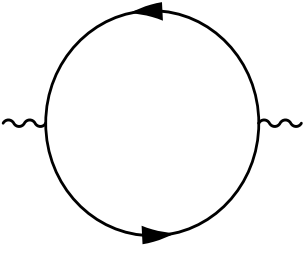} 
\caption{The one-loop boson self energy.}
\label{fig:RPA}
\end{figure}

\begin{figure}
        \includegraphics[height=3cm,width=4cm]{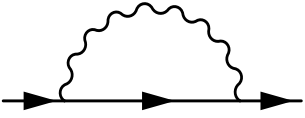} 
\caption{The one-loop fermion self energy.
Here the boson propagator is a dressed propagator 
which include the one-loop self energy correction in Fig. \ref{fig:RPA}. 
}
\label{fig:1loopfse}
\end{figure}

In the one-loop order, singular self energies are generated 
from the diagrams in Figs. \ref{fig:RPA} and \ref{fig:1loopfse},
and the quantum effective action becomes
\bqa
\Gamma & = &  \sum_{j} \int dk  
\Bigl[ i \frac{c}{N} ~\mbox{sgn}(k_0) |k_0|^{2/3} 
+  i k_0  \nn
&& ~~~~~~~~~~~~~~~ +  v_x k_x + v_y k_y^2 \Bigr] \psi_{j}^*(k) \psi_{j}(k)  \nn
&& + \int  dk 
 \left[ \gamma \frac{|k_0|}{|k_y|} +  k_0^2 + k_x^2 +  k_y^2 \right] a^*(k) a(k) \nn
&& + \frac{e}{\sqrt{N}} \sum_{j}  \int dk dq ~~  a(q) \psi^*_{j}(k+q) \psi_{j}(k),
\label{ga}
\eqa 
where $c$ and $\gamma$ are constants of the order of $1$.
To compute the fermion self energy,
the dressed boson propagator has been used
because the boson self energy is of the order of $1$.
In the low energy limit, 
the leading terms of the quantum effective action are invariant under the scale transformation,
\bqa
k_0 & = & b^{-1} k_0^{'}, \nn
k_x & = & b^{-2/3} k_x^{'}, \nn
k_y & = & b^{-1/3} k_y^{'}, \nn
\psi_{a}(b^{-1} k_0^{'}, b^{-2/3} k_x^{'}, b^{-1/3} k_y^{'}) & = & b^{4/3} 
\psi_{a}^{'}( k_0^{'},  k_x^{'}, k_y^{'}), \nn
a(b^{-1} k_0^{'}, b^{-2/3} k_x^{'}, b^{-1/3} k_y^{'}) & = & b^{4/3} 
a( k_0^{'},  k_x^{'}, k_y^{'}).
\label{scale3}
\eqa
The singular self energies render the terms,
\bqa
i k_0 \psi_{j}^*(k) \psi_{j}(k) , \nn
\left[ k_0^2 + k_x^2 \right] a^*(k) a(k) 
\eqa
irrelevant in the low energy limit.
Usually, it is expected that 
one restores the same low energy quantum effective action
if one drops the irrelevant terms from the beginning.
However, this is not true in this case.
If one drops the irrelevant terms in the bare action,
then the resulting theory
becomes completely localized in time 
and one can not have a propagating mode.
If there is no frequency dependence in the bare action,
the frequency dependent singular self energies can not be generated either.
Therefore, 
to restore the full low energy dynamics, one has to keep 
a minimal information that the theory is not completely localized in time.
It turns out that the following action given by
\bqa
{\cal L} & = & \sum_j \psi^*_{j} 
( \eta \partial_\tau - i v_x \partial_x - v_y \partial_y^2 ) \psi_{j} \nn
&& + \frac{e}{\sqrt{N}} \sum_j  a \psi^*_{j} \psi_{j} 
 + a ( - \partial_y^2 ) a,
\label{a3}
\eqa
is the minimal local theory which restores the 
one-loop quantum effective action (\ref{ga}).
Here $\eta$ is a parameter 
which has the dimension $-1/3$ according to the scaling (\ref{scale3}).

Since the time derivative term is irrelevant,
$\eta$ will flow to zero in the low energy limit, and
the bare value of $\eta$ does not affect any low energy physics 
as far as it is nonzero.
The role of the nonzero $\eta$ is to give 
a non-trivial frequency dependent dynamics
by maintaining the minimal causal structure of the theory
before it dies off in the low energy limit.
For example, 
in the computation of the one-loop 
boson self energy (Fig. \ref{fig:RPA}) in
\bw
\bqa
\Pi(q) & = & 
e^2  
\int d^3 k 
\frac{1}{  i \eta ( k_0 + q_0) +  v_x (k_x + q_x) + v_y (k_y + q_y)^2}~
\frac{1}{ i \eta  k_0 + v_x k_x + v_y k_y^2} = \gamma \frac{|q_0|}{|q_y|},
\label{eq:pi}
\eqa
\ew
the sign of $\eta$ contains the information on
whether the pole is on the upper or lower side 
in the complex plane 
for the $k_x$ integration.
The final result is independent of $\eta$.
As far as the `topological' information
on the location of poles is kept for the fermions, 
it generates the correct frequency dependent 
self energy for the boson in Eq. (\ref{eq:pi}).
Therefore we can completely drop the time derivative term of the boson
in the bare action (\ref{a1}).
The boson self energy, in turn, generates
the frequency dependent fermion self energy through Fig. \ref{fig:1loopfse}.

\subsection{Low energy limit and large $N$ limit}

At low energy $k_0 << E_\eta$ with $E_\eta = ( N \eta )^{-3}$, 
the dynamically generated fermion self energy is dominant
over the bare term $i \eta k_0$.
Here $E_\eta$ is a cross-over energy scale 
below which physics is described by the 
scale-invariant universal theory.
To study the low energy physics,
we will fix our energy scale $E$ and
send a UV cut-off $\Lambda$ and the 
cross-over scale $E_\eta$ to infinity\cite{UV}.
In taking the low energy limit,
it is convenient to maintain
the UV cut-off $\Lambda$ to be smaller than 
the cross-over scale,
that is, 
\bqa
E << \Lambda << E_\eta.
\label{eq:limit}
\eqa
First, a Feynman diagram with an external energy $E$ is computed 
with finite $\eta$, $\Lambda$ and $N$.
To maintain (\ref{eq:limit}),
we take the \EGZ limit first 
and then $\Lambda \rightarrow \infty$ limit later.
Finally, we take the large $N$ limit.
This amounts to imposing 
the condition (\ref{eq:limit}) for all $N$ 
as $N$ is progressively increased in the large $N$ limit.
In this way, we can keep the bare time derivative term 
to be always smaller compared to the singular self energy
at all energy scales.
In this limit,
not only the IR physics, but also 
the UV physics is controlled by
the same universal theory.
This is particularly convenient to study 
universal low energy dynamics of the theory
at the critical dimension which is $d_c = 2+1$
in this case.
This is because any logarithmic IR divergence
is reflected to a UV divergence,
and one can read the renormalization group flow 
by keeping track of UV divergences.
We will exploit this property 
to study dynamical properties
of the theory in Sec. IV.

The action (\ref{a3}) has four terms which are marginal at the one-loop level.
On the other hand, there are five parameters that set 
the scales of energy-momentum
and the fields.
Out of the five parameters, only four of them can modify 
the coefficients of the marginal terms
because the marginal terms remain invariant under 
the transformation (\ref{scale3}).
Using the remaining four parameters, 
one can always rescale 
the coefficients of the marginal terms
to arbitrary values.
Therefore, there is no dimensionless parameter in 
this theory except for the fermion flavor $N$.
In the following, we will set $v_x = v_y=e=1$.
With this choice, $c$ and $\gamma$ in Eq. (\ref{ga}) 
are automatically of the order of $1$.
The coefficients of the non-local terms 
are not independent tunable parameters because 
those parameters are completely determined from 
the local terms.

\section{$1/N$ expansion}

\subsection{Failure of a perturbative $1/N$ expansion}

In the naive counting of power in $1/N$,
a vertex contributes $N^{-1/2}$ 
and a fermion loop contributes $N^1$.
In this counting, 
only the fermion RPA diagram (Fig. \ref{fig:RPA}) 
is of the order of $1$,
and all other diagrams are of higher order in $1/N$.
In the leading order, the propagators become
\bqa
g_0(k) & = & \frac{1}{ i \eta k_0 +  k_x + k_y^2 } , \nn
D(k) & = & \frac{1}{ \gamma \frac{|k_0|}{|k_y|} +  k_y^2 }.
\label{propagators}
\eqa

\begin{figure}[h]
        \includegraphics[height=5cm,width=7cm]{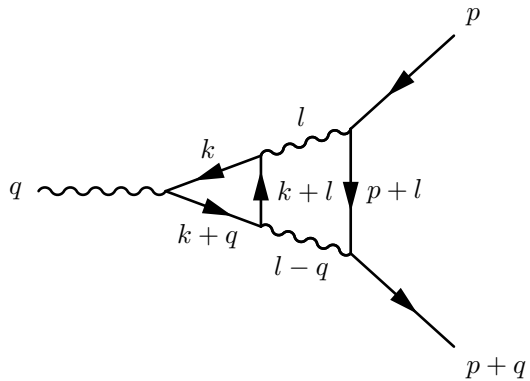} 
\caption{A two-loop vertex correction.}
\label{fig:vertex1}
\end{figure}

One can attempt to compute the full quantum effective action
by including $1/N$ corrections perturbatively.
However, we will see that this naive $1/N$ expansion
breaks down in the low energy limit.
To see this, let us consider a two-loop vertex correction
shown in Fig. \ref{fig:vertex1},
\bqa
\Gamma(p,p+q) & = &
- N^{-3/2} \int dk dl ~~~
g_0(k) g_0(k+q) \nn
&&  g_0(k+l) g_0(p+l) 
 D(l) D(l-q). 
\label{eq:vertex1}
\eqa
Let us focus on the case with $p=0$.
Without loss of generality, we can assume $q_0, q_y > 0$.
Integrating over $k_{x}$, $k_y$ and $l_x$, one obtains
\bqa
\Gamma(0,q) & =  &
- N^{-3/2} \int dl_0 dl_y dk_0  
\frac{F(l_0,l_y,k_0,q_0,q_y)}{ l_y \delta_{q}
+ i \eta l_y q_0 } \nn
\eqa
where
\bqa
&& F(l_0,l_y,k_0,q_0,q_y) = 
4 \pi^3 i 
\left[ \Theta(l_0+k_0) - \Theta(l_0) \right] \times \nn
&& ~~~~ \left[ \Theta(k_0+q_0) - \Theta(k_0) \right]
\left[ \Theta(q_y-l_y) - \Theta(q_y) \right] D(l) D(l-q) \nn
\eqa
is a function 
which is independent of $\eta$ and $N$, 
with $\Theta(x)$ being a step function,
and $\delta_q = q_x + q_y^2$ is the `distance' of ${\bf q}$ 
from the Fermi surface.
If the final momentum of the fermion is also on the Fermi surface,
that is, $\delta_{q}=0$,
the vertex correction becomes
\bqa
\Gamma(0,q) = -\frac{N^{-3/2}}{\eta q_0^{1/3}} f_1( q_y/q_0^{1/3} ),
\label{eq:vertex2}
\eqa
where $f_1(t)$ is a non-singular universal function 
which is independent of $N$ and $\eta$,
\bqa
f_1(t) &=& 4 \pi^3 \int_{-1}^0 dx \int_0^{|x|} dy \int_1^\infty dz  \nn
&& \frac{ t^2 (z-1) }{ (\gamma y + (tz)^3 ) ( \gamma (1-y) + t^3(z-1)^3 )}.
\eqa
In the \EGZ limit, this two-loop vertex correction 
which connects two fermions on the Fermi surface
diverges.
This divergence is quite generic :
a vertex function which connects fermions on the Fermi surface
diverges as $1/\eta^n$ for some integer $n$ in general.
The physical reason for this divergence is simple.
In the \EGZ  limit, 
the bare fermion propagator is independent of frequency
and the integration over frequencies is ill-defined.
This divergence is unphysical 
in the sense that it disappears
once the frequency dependent fermion self energy correction is included.
If one include the one-loop fermion self energy (Fig. \ref{fig:1loopfse}),
the dressed fermion propagator becomes
\bqa
g(k) = \frac{1}{ 
i \eta k_0 + i \frac{c}{N} ~\mbox{sgn}(k_0) |k_0|^{2/3} 
+ k_x + k_y^2 },
\eqa
and the $1/\eta$ divergence disappears.
Instead, the resulting finite term becomes
enhanced by a factor of $N^n$ 
for some integer $n \geq 0$, 
because the zero  in the denominator (in the \EGZ limit)
is replaced by a term 
which is proportional to $1/N$.
As a result, the two-loop vertex correction 
shown in Fig. \ref{fig:vertex1}
becomes 
\bqa
\Gamma(0,q) = - N^{-1/2} f_2(q_y/q_0^{1/3} ),
\label{eq:vertex3}
\eqa 
where $f_2(t)$ is a non-singular universal function 
which is independent of $N$ and $\eta$,
\bw
\bqa
f_2(t)  =  \frac{4 \pi^3}{c} \int_{-1}^0 dx \int_0^{|x|} dy \int_1^\infty dz  
&& 
\frac{1}{ | x+1|^{2/3} +  y^{2/3} + | x+y|^{2/3} + (z-1)( |x+1|^{2/3} + |x|^{2/3} ) } \times \nn 
&&  \frac{ t^2 z (z-1) }{ (\gamma y + (tz)^3 ) ( \gamma (1-y) + t^3(z-1)^3 )}.
\eqa
\ew
The additional factor of $N$ is 
from the enhancement factor
that arises due to the $1/\eta$ divergence
when the fermions are on the Fermi surface.
With the inclusion of the fermion self energy,
the IR divergence in Eq. (\ref{eq:vertex2})
has been traded with an enhanced power in $N$
in Eq. (\ref{eq:vertex3}).

\begin{figure}[h]
        \includegraphics[height=4cm,width=5cm]{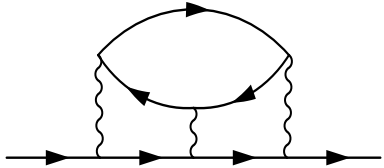} 
\caption{A three-loop fermion self energy correction.}
\label{fig:fse1}
\end{figure}

Similar enhancement factors arise in other diagrams as well.
For example, a three-loop fermion self energy correction
shown in Fig. \ref{fig:fse1} is of the order of $N^{-2}$
according to the naive counting.
However, the self energy 
of fermion on the Fermi surface ($\delta_p=0$) 
diverges as $1/\eta$ in the \EGZ limit 
if the bare fermion propagator in Eq. (\ref{propagators}) is used.
If one includes the one-loop self energy of fermion,
it becomes of the order of $N^{-1}$,
\bqa
\Sigma(p) = - i \frac{c_3}{N} \mbox{sgn}(p_0) |p_0|^{2/3},
\eqa 
when the external fermion is on the Fermi surface.
Here $c_3$ is a universal constant of the order of $1$. 

This discrepancy 
between the cases with a finite $\eta$ and 
an infinitesimally small $\eta$ 
can be understood is the following way.
With a finite $\eta$, 
there is a crossover around the scale $q_0 \sim E_\eta$.
For $q_0 >> E_\eta$,
the $i \eta k_0$ is dominant in the fermion propagator
and a Feynman diagram obeys the naive counting in $1/N$.
On the other hand, for $q_0 << E_\eta$
quantum fluctuations are controlled by
the non-local term which is suppressed by $1/N$.
The enhanced quantum fluctuations at low energies 
lead to an enhancement factor by a positive power in $N$.
Since we are concerned about the low energy physics,
we should consider the latter limit.
This correct low energy limit is automatically taken by considering
the \EGZ limit with a fixed energy scale $q_0$.
This enhancement in the power of $N$ at IR 
is a manifestation of the fact
that quantum fluctuations become stronger at low energies.

\subsection{Genus Expansion}

In the low energy limit,
what determines the power of a Feynman diagram 
in $1/N$ ?
To answer this question, one should 
understand the origin of the 
enhancement factor 
discussed in the previous section more systematically. 
In the present section, 
we will develop a simple geometrical
way of determining power of
general Feynman graphs.

First, we illustrate the basic idea 
using the example (Fig. \ref{fig:vertex1}) considered in the previous section.
As we have seen in the previous section,
the enhancement factor $N$
is a consequence of the $1/\eta$ singularity 
in the \EGZ limit.
To understand the origin of the $1/\eta$ singularity, 
it is useful to examine the way
fermions are scattered near the Fermi surface.
Suppose both ${\bf p}$ and ${\bf p}+{\bf q}$ are on the Fermi surface in Fig. \ref{fig:vertex1}.
In the fermion loop with running momentum $k$,
the momentum of the fermion consecutively becomes
$k, (k+q), (k+l)$
as a result of scatterings.
For a given external momentum $q$ of the boson,
one can always choose 
the spatial momentum ${\bf k}$ 
to make both ${\bf k}$ and ${\bf k}+{\bf q}$ to 
be on the Fermi surface.
There is one unique choice, ${\bf k} = {\bf p}$.
To make the next momentum ${\bf k}+{\bf l}$ 
to be on the Fermi surface as well,
one needs to tune only $l_x$
and there is one remaining free parameter, 
${l}_{y}$.
This is because the Fermi surface
has dimension one.
As a result, all internal fermions can remain 
right on the Fermi surface during the 
scattering process 
no matter what the
values of one momentum component 
($l_{y}$) is
if the other three momentum components 
($k_{x}$, $k_{y}$, $l_{x}$) are finely tuned.
This implies that 
all four fermion propagators are singular at $\eta=0$ 
in an one-dimensional manifold 
which is embedded in the four dimensional space 
${\bf k}, {\bf l}$.
We refer to this manifold as
a `singular manifold'.
The co-dimension of the singular manifold
is $4 - 1 = 3$,
and there are only $3$-dimensional integrations
which contribute to the phase space volume
and cancel the IR divergence.
Since the product of the propagators has a singularity
whose strength is $4$,
the integration over the $3$ parameters
can not completely remove the singularity
and the remaining singularity becomes of the order of 
$4-3 = 1$.
This explains 
why Fig. \ref{fig:vertex1} 
has the singularity of $1/\eta$ 
when the bare fermion propagator is used,
and why it has the enhancement factor $N$
when the one-loop dressed propagator  is used.
The enhancement factor $N$ for the fermion self energy 
in Fig. \ref{fig:fse1} 
can be understood in the similar way.

\begin{figure}
        \includegraphics[height=3.5cm,width=8cm]{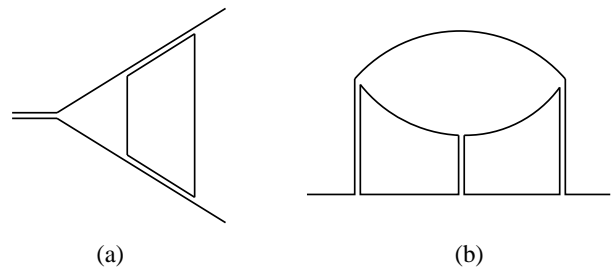} 
\caption{The double line representations of 
Fig. \ref{fig:vertex1} (a) and Fig. \ref{fig:fse1} (b).
Double lines represent propagators of the boson, 
and the single lines are the propagators of the fermion.
The number of single line loops (one in (a) and two in (b)) 
represents the dimension of the singular manifold (see the text) 
on which all fermions remain on the Fermi surface
in the space of internal momenta. 
}
\label{fig:vertex_fse_double}
\end{figure}

What determines the dimension of the singular manifold
within which fermions always remain on the Fermi surface?
It turns out that the dimension of 
the singular manifold is given by
the number of closed loops 
when one draws boson propagators
using double lines 
and fermion propagators 
using single lines.
This can be shown by following 
the scheme used to prove the Migdal's theorem in 
the electron-phonon system\cite{SHANKAR,TSAI}.
First, we restrict momenta of all fermions to be on the Fermi surface.
A momentum ${\bf k}_\theta$ of fermion on the Fermi surface
is represented by an one-dimensional parameter $\theta$.
Then, a momentum of the boson $\textbf{q}$
is decomposed into two momenta on the Fermi surface
as ${\bf q}={\bf k}_\theta-\textbf{k}_{\theta^{'}}$,
where both $\textbf{k}_\theta$ and $\textbf{k}_{\theta^{'}}$ 
are on the Fermi surface.
This decomposition is unique
because there is only one way of choosing 
such $\textbf{k}_\theta$ and $\textbf{k}_{\theta^{'}}$ near $\textbf{k}=0$.
As far as momentum conservation is concerned,
one can view the boson of momentum $\textbf{q}$
as a composite particle made of 
a fermion of momentum $\textbf{k}_\theta$
and a hole of momentum $\textbf{k}_{\theta^{'}}$.
For example, the two-loop vertex correction in Fig. \ref{fig:vertex1}
and the three-loop fermion self energy correction in Fig. \ref{fig:fse1}
can be drawn as Fig. \ref{fig:vertex_fse_double} (a) and (b) respectively
in this double line representation.
In this representation, each single line
represents a momentum on the Fermi surface.
Momenta in the single lines that are connected 
to the external lines should be uniquely fixed
to make all fermions stay on the Fermi surface.
On the other hand, momenta on 
the single lines that form closed loops
by themselves are unfixed.
In other words, all fermions can stay on the Fermi surface
no matter what the value of the unfixed momentum component
that runs through the closed loop is.
Since there is one closed loop in Fig. \ref{fig:vertex_fse_double} (a),
the dimension of the singular manifold is $1$
and the enhancement factor becomes $N^{4-(4-1)}=N$
for the two-loop vertex correction in Fig. \ref{fig:vertex1}.
In Fig. \ref{fig:vertex_fse_double} (b), there are two closed loop,
and the enhancement factor for Fig. \ref{fig:fse1} 
becomes $N^{5-(6-2)}=N$ : 
there are five fermion propagators,
six spatial components of internal momenta
and two closed loops.

The enhancement factor is
a direct consequence of the presence of 
infinitely many soft modes associated with
deformations of the Fermi surface.
The extended Fermi surface makes it possible
for virtual particle-hole excitations
to maneuver on the Fermi surface
without costing much energy.
As a result, quantum fluctuations becomes 
strong when external momenta are arranged 
in such a way that there are sufficiently 
many channels for the virtual particle-hole 
excitations to remain on the Fermi surface.
This makes higher order processes to be 
important even in the large $N$ limit.
We note that this effect is absent in relativistic quantum field theories 
where gapless modes exist only at discrete points 
in the momentum space.

Now we are ready to write down a general formula
which tells order of a general Feynman diagram in $1/N$. 
\begin{itemize}
\item
First, draw a Feynman diagram 
using single lines for fermion propagators
and using double lines for boson propagators.

\item
Second, each vertex contributes $1/\sqrt{N}$.

\item
Third, each fermion loop contributes $N$.

\item
Finally, the enhancement factor is given by
\bqa
N^{[ I_f - 2L + n]}.
\eqa
Here $I_f$ is the number of (internal) fermion propagators,
$L$ is the number of loops (the number of internal momenta)
and $n$ is the number of closed single line loops
in the double line representation.
$[x]=x$ if $x \geq 0$
and $[x]=0$ if $x<0$.
This enhancement factor can be understood as 
was illustrated in the previous examples.
When all fermions are on Fermi surface,
the product of propagators has 
the singularity of the order of $I_f$.
Upon integrating over the internal momenta,
the singularity is lowered
due to the contribution from phase space volume.
There are $2L$ components of spatial momenta, 
but $n$ of them are degenerate in that
all fermions remain on the Fermi surface
no matter what the values of the $n$ momentum components are
as far as the remaining $2L-n$ components are zero.
Therefore, the integration over the internal momenta
can remove the singularity 
only by the power of $(2L-n)$.
The power of the remaining IR divergence is $I_f - (2L-n)$
and this results in the enhancement factor, $N^{(I_f - 2L + n)}$.
If $(2L-n) > I_f$, 
the suppression from the phase space of internal momenta is more than 
enough to suppress the whole singularity, 
and there is no
enhancement factor.
In this case, the enhancement factor should be $1$, 
not a negative power of $N$.
That is why we use $[I_f-2L+n]$ which is $0$ if $I_f<(2L-n)$.
By using the relation between $L$ and $I_f$,
$I_f = 2L + \frac{E_f+2 E_b}{2}-2$,
where $E_f$ ($E_b$) is the number of external fermion (boson) lines,
one can write the enhancement factor as
\bqa
N^{\left[n + \frac{E_f+2 E_b}{2}-2 \right]}.
\eqa

\end{itemize}

As a result, the net order of a Feynman diagram 
is given by $N^Q$ with
\bqa
Q = - \frac{V}{2} + L_f + \left[n + \frac{E_f+2 E_b}{2}-2 \right],
\label{power}
\eqa
where $V$ is the number of vertices
and $L_f$ is the number of fermion loops.

\begin{figure}[h]
        \includegraphics[height=4.5cm,width=5cm]{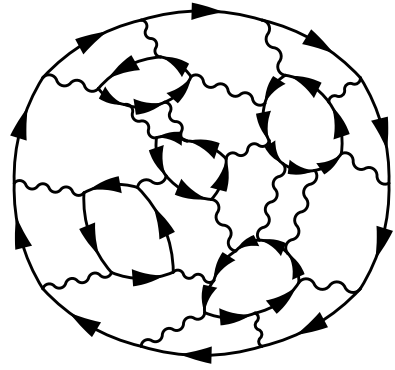} 
\caption{A typical vacuum planar diagram which is of the order of $N^0$.
In planar diagrams, all fermion propagators which face to each other
flow in the opposite direction.
In this way, fermions can stay on the Fermi surface before and after scatterings.
}
\label{fig:vacuum_planar}
\end{figure}

\begin{figure}
        \includegraphics[height=5.0cm,width=5.0cm,angle=-30]{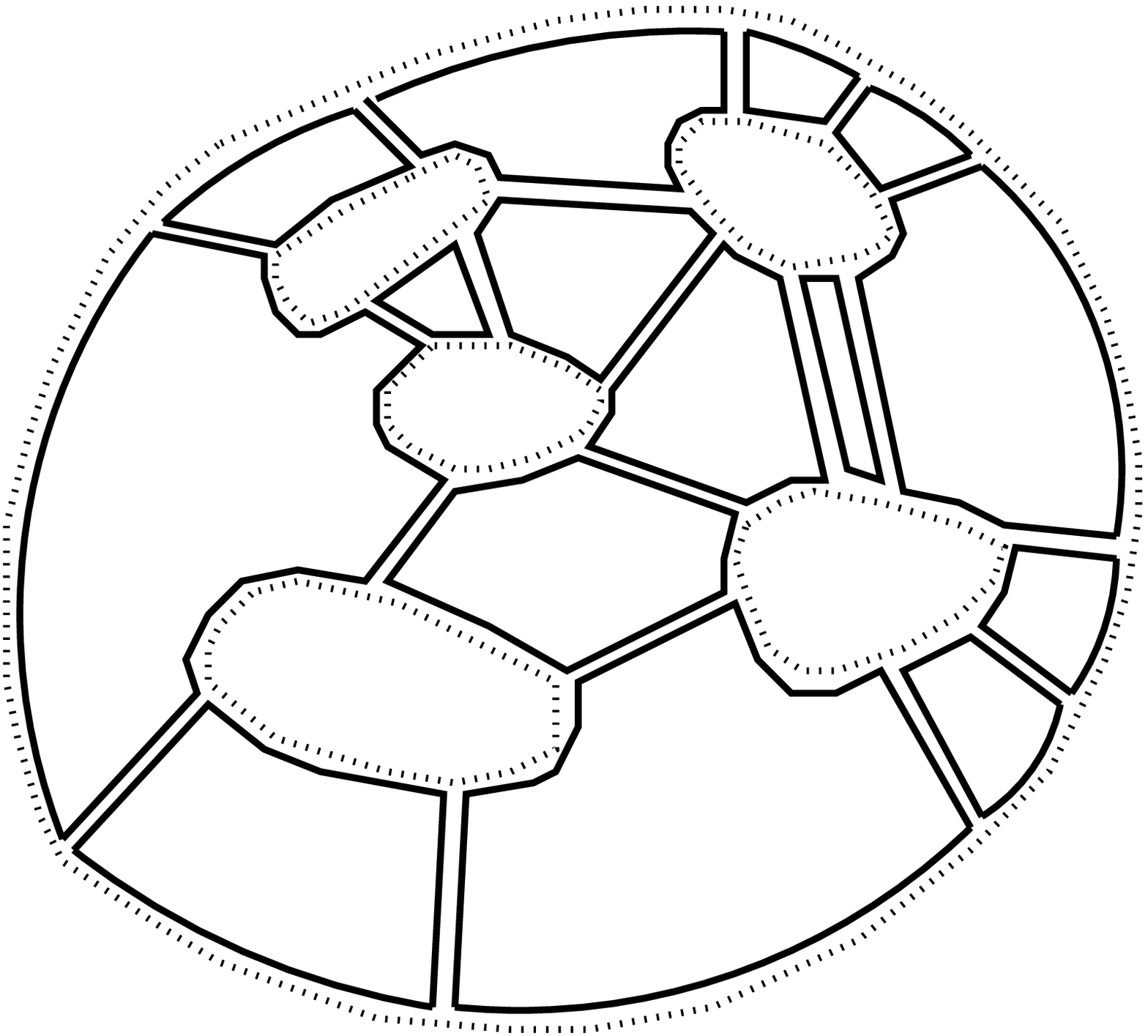} 
\caption{The full double line representation of 
the planar diagram shown in Fig. \ref{fig:vacuum_planar}.
One can draw this diagram on the sphere without any crossing.
The solid double lines represent the boson propagator
and double lines made of one solid and one dotted lines 
represent fermion propagators.
Loops of dotted lines are added to each fermion loops.
In this representation, there is a factor of $N$
for each closed single line loop 
whether it is a loop made of a solid or dotted line.
}
\label{fig:planar_double_double}
\end{figure}

Now let us classify Feynman diagrams 
based on the expression Eq. (\ref{power}),
starting from vacuum diagrams.
Classification of non-vacuum diagrams with external lines 
naturally follows from that of vacuum diagrams, as will be shown shortly.
The leading order vacuum diagram is 
the one fermion loop diagram 
which is of the order of $N$.
In the next order of $N^{0}$, 
there are infinitely many diagrams.
A typical diagram of the order of $N^0$ 
is shown in Fig. \ref{fig:vacuum_planar}.
For the diagram in Fig. \ref{fig:vacuum_planar}, 
we have
$V=38$, $n=15$, $E_f=E_g=0$ and $L_f=6$, which gives
\bqa
Q = -19 + 6 + [15-2] = 0.
\eqa
Actually, there is a simple geometrical way of interpreting the result.
First, we turn fermion propagators into double lines as well
by drawing additional single line loops 
for each fermion loop as in Fig. \ref{fig:planar_double_double}.
In this way, we can include the factor $N^{L_f}$ 
from fermion loops by counting the additional 
closed loops of single lines.
We will refer to this way of drawing a `full double line representation'.
If $n \geq 2$, 
which is always the case for sufficiently large $V$ 
if there are not too many crossings, 
we can remove the bracket in Eq. (\ref{power})
and the power can be rewritten as
\bqa
Q = V - I + F -2.
\eqa
Here we use the identity $3V = 2I$, where
$I$ is the number of total internal propagators,
$F = n + L_f$ is the total number of single line loops 
including the additional single line loops added to 
each fermion loop.
In this full double line representation, 
one can think of 
a closed 2d surface 
formed by joining the patches of single line loops.
The 2d surface is the surface 
on which a Feynman diagram can be drawn
without any crossing in the
full double line representation.
The factor $\chi = V-I+F$ is nothing but the Euler number
of the 2d closed surface
and the power $Q$ becomes
\bqa
Q = -2 g,
\label{power1}
\eqa
where $g$ is the genus of the 2d surface.
The diagrams of the $N^0$ order are the planar diagrams
which can be drawn on a sphere.

\begin{figure}[h]
        \includegraphics[height=4.5cm,width=5cm]{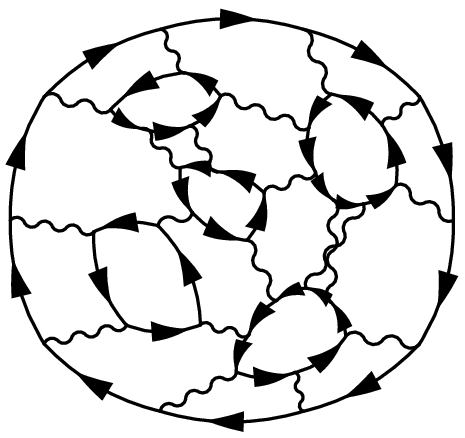} 
\caption{A non-planar diagram which is of the order of $N^{-2}$.}
\label{fig:vacuum_nonplanar}
\end{figure}

\begin{figure}
        \includegraphics[height=5.0cm,width=5.0cm,angle=-30]{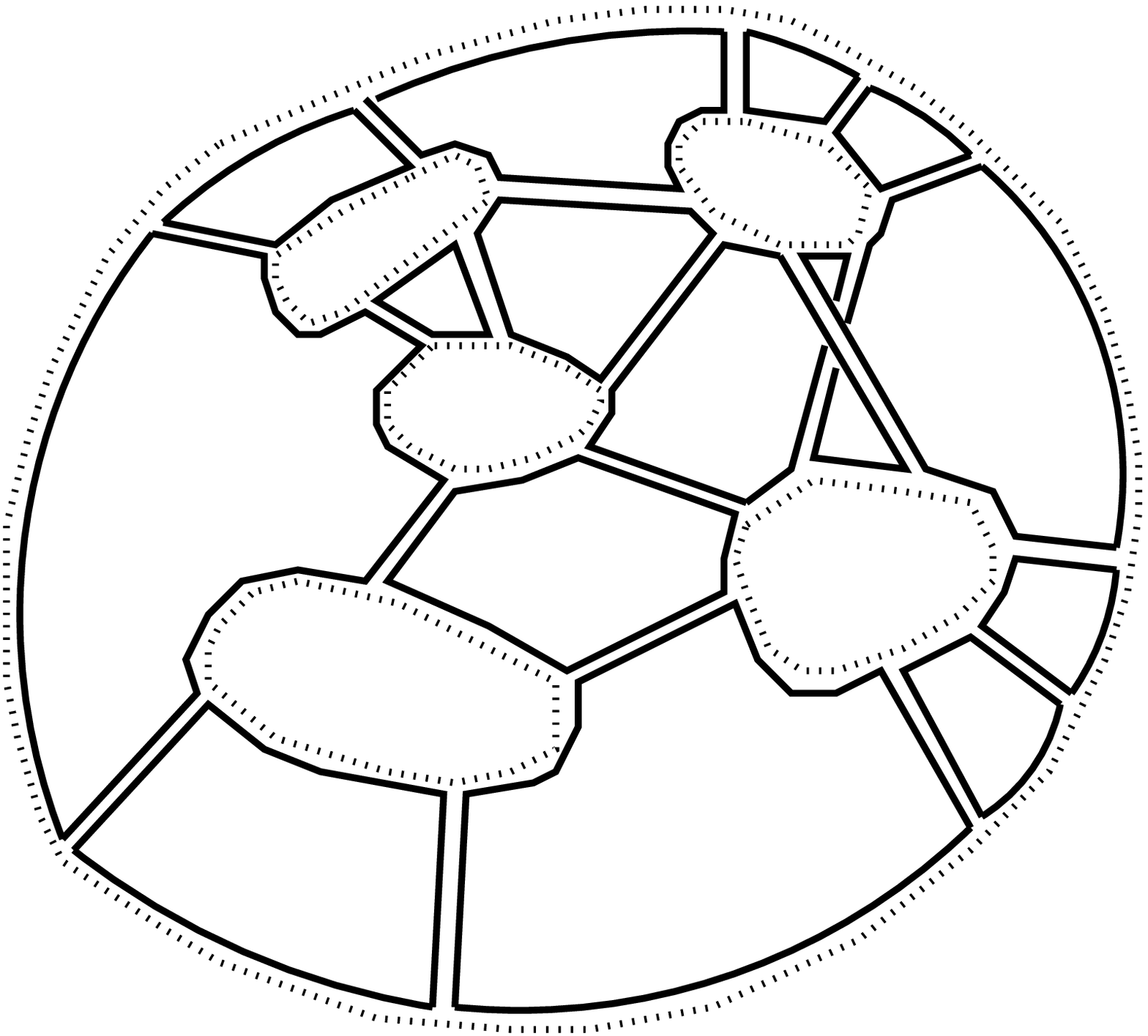} 
\caption{ 
The full double line representation of Fig. \ref{fig:vacuum_nonplanar}.
This diagram needs to be drawn on the surface of a torus to avoid a crossing.
}
\label{fig:nonplanar_double_double}
\end{figure}

For non-planar diagrams,
 such as the one shown in Fig. \ref{fig:vacuum_nonplanar},
one has to introduce closed surfaces with handles
to draw them in the full double line representation 
without a crossing.
This is illustrated in Fig. \ref{fig:nonplanar_double_double}.
Contributions from non-planar diagrams
are suppressed as the number of genus increases
according to Eq. (\ref{power1}).

\begin{figure}
        \includegraphics[height=5.0cm,width=5.0cm,angle=-30]{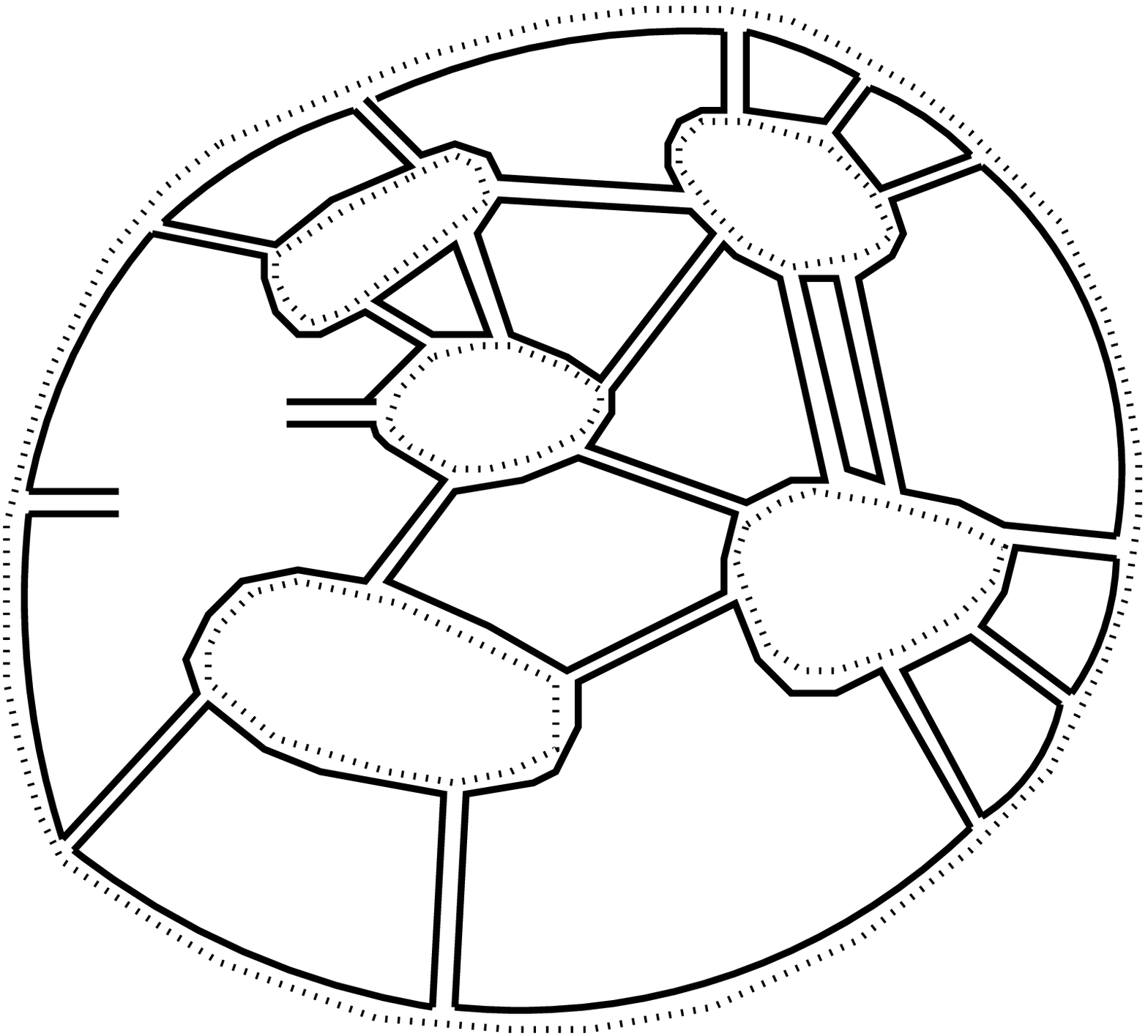} 
\caption{
A boson self energy diagram drawn in the full double line representation. 
It has been created by attaching two vertices to the vacuum diagram in Fig. \ref{fig:planar_double_double}.
This diagram is nominally of the order of $N^0$ for any external momentum.
But, it turns out that all planar boson self energy diagrams vanish (see the text).
}
\label{fig:boson_planar_double}
\end{figure}

\begin{figure}
        \includegraphics[height=5.0cm,width=5.0cm,angle=-30]{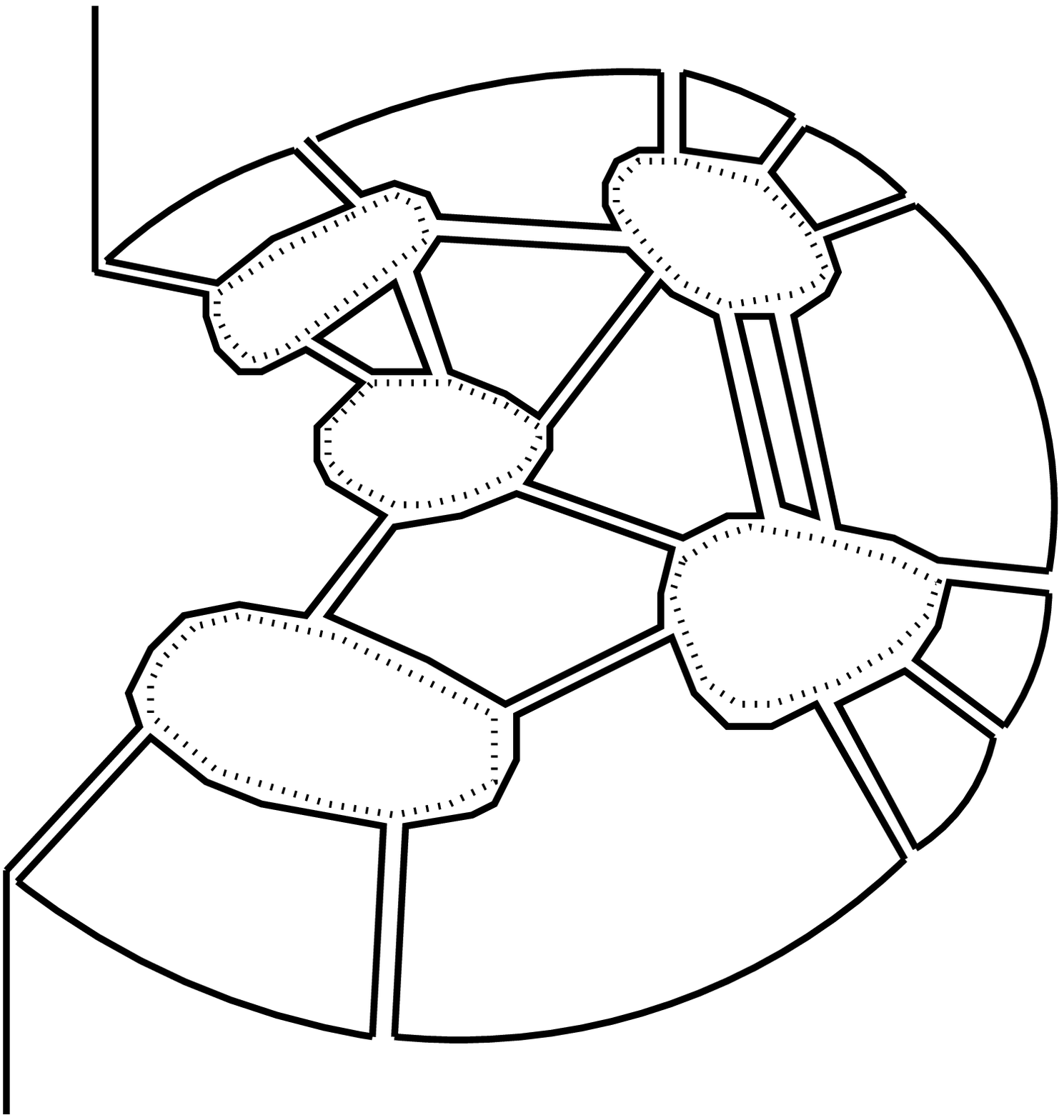} 
\caption{ 
A fermion self energy diagram created by cutting fermion propagator open in the vacuum diagram in Fig. \ref{fig:planar_double_double}.
This diagram is of the order of $N^{-1}$ when external momentum is on the Fermi surface. 
}
\label{fig:fermion_planar_double}
\end{figure}

\begin{figure}
        \includegraphics[height=5.0cm,width=5.0cm,angle=-30]{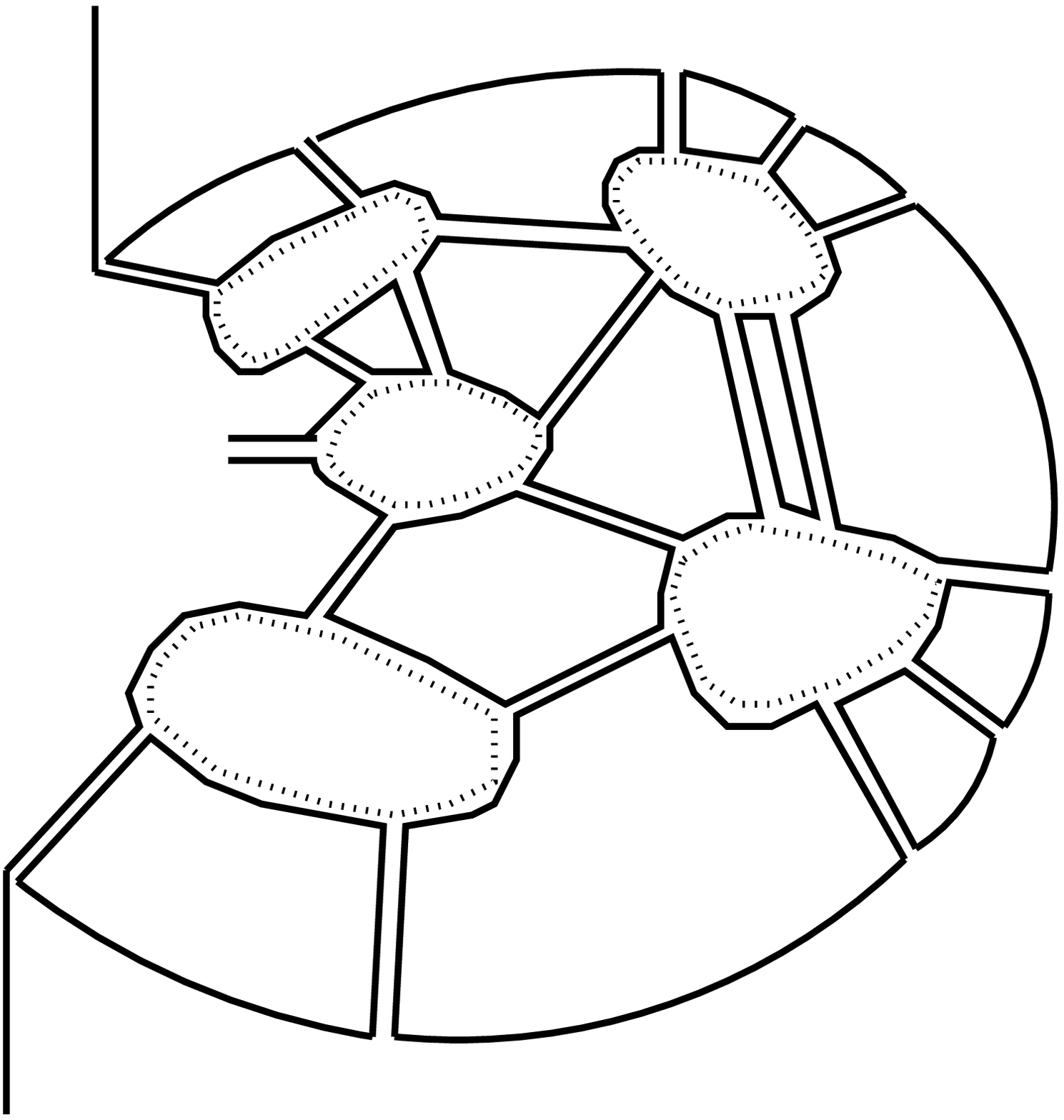} 
\caption{ 
A vertex correction created by cutting a fermion propagator open and attaching a vertex to the vacuum diagram in Fig. \ref{fig:planar_double_double}.
This diagram is of the order of $N^{-1/2}$ when external fermion is on the Fermi surface. 
}
\label{fig:vertex_planar_double}
\end{figure}

Power counting of diagrams with external lines
can be easily obtained from the counting of vacuum diagrams.
To create a boson self energy diagram,
one attaches two vertices to fermion propagators.
The leading order self energy diagrams can be generated
if two vertices are attached to fermion propagators
which are parts of one single line loop.
A typical leading order boson self energy diagram created from a planar
vacuum diagram is shown in Fig. \ref{fig:boson_planar_double}.
From this procedure, the boson self energy diagram
acquires the additional power of $N^{-1}$ from two added vertices ($\Delta V=2$),
$N^{2}$ from two external boson lines ($\Delta E_b = 2$)
and $N^{-1}$ from a lost single line closed loop ($\Delta n = -1$).
As a result, the resulting boson self energy diagram
has the same power as the parent vacuum diagram
which is of the order of $N^{0}$
for planar diagrams.
It is noted that the one-loop boson self energy diagram (Fig. \ref{fig:RPA}),
which can be created by attaching two vertices to 
the vacuum diagram of the order of $N$, is also order of $N^0$.
Therefore there are infinitely many planar diagrams which contribute
to the boson self energy in the leading order of $N^0$.
If one attaches two vertices to two fermion propagators
which belong to different single line loops, or
if one starts from a non-planar vacuum diagram,
the resulting boson self energy is down by an additional factor of $1/N$.

Fermion self energy diagrams can be created by
cutting a fermion propagator
as in Fig. \ref{fig:fermion_planar_double}.
This procedure causes 
an additional power of $N^{-1}$ from one less
single line loop ($\Delta n = -1$),
$N^{1}$ from two external fermion lines
($\Delta E_f = 2$), and 
$N^{-1}$ from one less fermion loop ($\Delta L_f = -1$).
As a result, the fermion self energy diagram is down by $N^{-1}$ 
from the vacuum diagram.
Therefore the leading order fermion self energy corrections 
are of the order of $N^{-1}$.
There are infinitely many planar diagrams that contribute
to the leading frequency dependence of the fermion propagator
which is of the order of $N^{-1}$.

Leading order three-point vertex functions
can be created by cutting a fermion propagator
and attaching a vertex to another fermion propagator
as in Fig. \ref{fig:vertex_planar_double}.
The resulting diagram is of the order of $N^{-1/2}$ :
an additional power of 
$N^{-1/2}$ from $\Delta V = -1$,
$N^{-1}$ from $\Delta n = -1$,
$N^{2}$ from $\Delta E_f = 2$ and $\Delta E_b = 1$, and 
$N^{-1}$ from $\Delta L_f = -1$.
All planar vertex corrections are 
of the same order as the bare vertex.

For the fermion self energy
and the 3-point vertex function,
the above counting is valid only when
the external fermion momenta are sufficiently
close to the Fermi surface.
If external momenta are far from the Fermi surface,
one can not make all internal fermions to 
be on the Fermi surface without 
an additional tuning of internal momenta.
As a result, the enhancement factor becomes smaller
than what is predicted for the case when 
external fermions are on the Fermi surface.
To be more precise, Eq. (\ref{power}) is valid if 
\bqa
|\delta_k|  << \frac{|k_0|^{2/3}}{N},
\eqa
where $k$ is the momentum of the external fermion.
In the opposite limit, 
$|\delta_k|  >> \frac{|k_0|^{2/3}}{N}$,
there are additional factors in $1/N$.

On the other hand, for the boson self energy,
Eq. (\ref{power}) is always valid,
irrespective of the energy and momentum of
the external boson.
This is because any boson momentum can be decomposed
into two momenta on the Fermi surface and 
one can use the double line representation,
which leads to the counting in Eq. (\ref{power}).

\begin{figure}[h]
        \includegraphics[height=5cm,width=6cm]{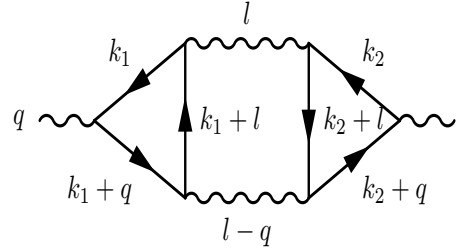} 
\caption{A planar three-loop boson self energy.}
\label{fig:ladder1}
\end{figure}

The leading contributions come from the 
planar diagrams where the genus of the underlying
2d surface is zero.
In principle, there can be infinitely many diagrams
which are order of $N^0$ ($N^{-1}$) for the boson 
(fermion) self energy and $N^{-1/2}$ 
for the three-point vertex function.
In particular, the boson self energy has infinitely many
leading order terms at any external momentum.
This would have made the one-loop 
boson propagator unreliable
even in the large $N$ limit.
However, it turns out that all 
planar diagrams for the boson self energy correction
beyond the one-loop level vanish
due to a kinematical reason.
To see this,
let us consider a three-loop boson self energy correction
shown in Fig. \ref{fig:ladder1},
\bqa
\Pi(q) & = & - N^{-1} \int dk_1 dk_2 dl~~
g(k_1)
g(k_1+q)
g(k_1+l) \nn
&& g(k_2)
g(k_2+q)
g(k_2+l)
D(l)
D(1-q).
\label{eq:ladder1}
\eqa
This diagram is nominally of the order of $N^0$ 
due to an enhancement factor $N^1$.
Integrating over $\delta_l$, one obtains
\bqa
&& \Pi(q)  \sim 
 \int d \delta_{k_1} d \delta_{k_2} d_{k_{1y}} d_{k_{2y}} dk_{10} dk_{20} 
dl_{0} dl_{y} ~ D(l) D(l-q) \times \nn
&&
\frac{1}{ \delta_{k_1} + i \{ k_{10} \} }
\frac{1}{ \delta_{k_1} + \delta_{q} + 2 k_{1y} q_y + i \{ k_{10} + q_0\} } \times \nn
&&
\frac{1}{ \delta_{k_2} + i \{ k_{20} \} }
\frac{1}{ \delta_{k_2} + \delta_{q} + 2 k_{2y} q_y + i \{ k_{20} + q_0\} } \times \nn
&&
\frac{ \Theta(k_{10}+l_{0}) - \Theta(k_{20}+l_{0}) }
{ \delta_{k_1} - \delta_{k_2}  + 2(k_{1y}-k_{2y})l_{y} 
+ i ( \{ k_{10} + l_0 \} - \{  k_{20} + l_0 \} )}, \nn 
\eqa
with 
$\delta_k = k_x + k_y^2$ and 
$\{ x \} =  \frac{c}{N} x |x|^{-1/3}$.
Now we change the integration variables as 
$k_{1y} = k$
and $k_{2y} = k + k^{'}$.
The integration over $k$ has poles 
at 
$i \{ k_{10} + q_0\} / 2q_y$
and
$i \{ k_{20} + q_0\} / 2q_y$,
and these poles are on the same side on the complex plane.
This can be seen from the fact that 
\bqa
|q_0| > \mbox{max}( |k_{10}|, |k_{20}|, |l_0| ).
\label{qmax}
\eqa
If (\ref{qmax}) is not true, one of 
$|k_{10}|, |k_{20}|, |l_0|$ should be the largest
among all frequencies.
If $|k_{10}|$ is the largest, then the integration 
over $\delta_{k_{1}}$ vanishes because all terms dependent on $\delta_{k_{1}}$
have poles on the same side in the complex plane.
This same argument applies to all other frequencies.
Therefore (\ref{qmax}) should be satisfied in order for
the integrations for $\delta_{k_{1}}$, $\delta_{k_{2}}$ and $\delta_l$ not to vanish.
Then
$i \{ k_{10} + q_0\} / 2q_y$
and
$i \{ k_{20} + q_0\} / 2q_y$ 
are on the same side in the complex plane,
and the integration over $k$ vanishes.
This proves that the Eq. (\ref{eq:ladder1}) vanishes.

\begin{figure}
        \includegraphics[height=6cm,width=7cm]{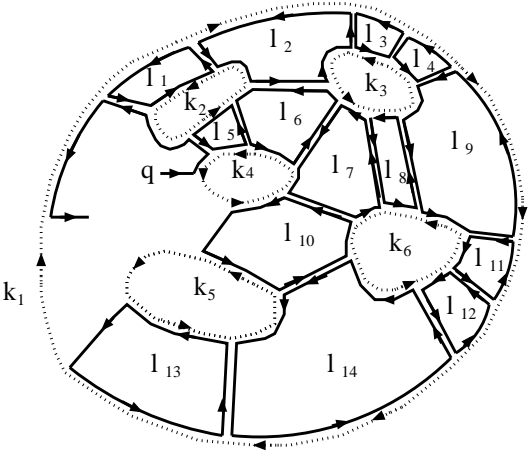} 
\caption{ 
The assignment of one internal momentum to each closed single line loop in the boson self energy diagram in Fig. \ref{fig:boson_planar_double}.
}
\label{fig:boson_planar_double_double_momenta}
\end{figure}

One can easily generalize the previous argument to prove
that all planar diagrams for boson self energy correction vanish.
Consider a general boson self energy diagram shown in Fig. \ref{fig:boson_planar_double}.
One can arrange internal momenta so that one internal momentum runs
within each single line loop as is shown in Fig. \ref{fig:boson_planar_double_double_momenta}.
Here $k_i$'s are momenta that run within fermion loops,
and $l_i$'s are momenta that connects different fermion loops through 
boson propagators.
If one performs the integrations over $\delta_{l_{i}}$, 
and changes the integration variables as
$\delta_{k_i} = k_{ix} + k_{iy}^2$ for all $i$,
$k_{1y} = k$,
$k_{2y} = k + k_2^{'}$,
$k_{3y} = k + k_3^{'}$, ...,
$k_{6y} = k + k_6^{'}$,
one can see that $k$ dependencies arise only from $g(k_i+q)$'s.
In order for the integrations for $\delta_{k_{i}}$ and $\delta_{l_{i}}$ 
not to vanish,
the external frequency should be the largest of all frequencies
in magnitude, that is,
$|q_0| > \mbox{max} ( |k_{i0}|, |l_{i0}|)$.
As a result, the integration over $k$ 
has all poles on one side in the complex plane :
all $q_0 + k_{i0}$ have the same sign as $q_0$.
Therefore all planar diagrams for boson self energy vanish.
The reader may wonder why the one-loop boson self energy 
(Fig. \ref{fig:RPA}) does not vanish.
This is because the integration over $k_y$ obtained after performing
the $k_x$ integration in Eq. (\ref{eq:pi}) has only one pole.
This is special for the one-loop diagram, and 
all higher-loop planar boson self energy graphs vanish.

Although the one-loop result is accurate for 
the boson self energy in the large $N$ limit,
there are infinitely many non-vanishing planar diagrams 
for fermion self energy and vertex correction.
This is because in those diagrams 
there exist `isolated' internal fermion propagators 
which are not part of any fermion loops 
(for example the fermion propagator 
with momentum $p+l$ in Fig. \ref{fig:vertex1}),
and the integration over $k$ (the uniform component of $k_{iy}$)
can have poles in both sides in the complex plane
due to contributions from the isolated fermion propagators.
All planar fermion self energy diagrams
are of the order of $N^{-1}$, and 
planar vertex corrections are order of $N^{-1/2}$,
when external fermions are on the Fermi surface.
Since the leading frequency dependence of the fermion propagator
is of the order of $N^{-1}$, one has to sum over 
all planar diagrams to extract 
dynamical properties of low energy fermions.
We note that the set of all planar diagrams 
is much larger than the set of rainbow diagrams 
that can be summed in a closed Dyson equation\cite{ALTSHULER}.
The fact that the low energy dynamics of fermion
is governed by the infinitely many planar diagrams implies 
that  
{\it 
fermions on the Fermi surface
remain strongly coupled even in the large $N$ limit.}

A few comments are in order for Eq. (\ref{power}).
First, the counting of power in $1/N$ is
self consistent in that we obtained Eq. (\ref{power})
based on the assumption (suggested by the one-loop result)
that the leading self energy corrections
of boson and fermion are of the order of $N^0$ and $N^{-1}$ respectively.
The fact that Eq. (\ref{power}) predicts the same conclusion
implies that the power counting will not change even though 
one uses the full propagators obtained by summing over all planar diagrams.
Second, some non-planar diagrams may have 
a higher power in $1/N$ than nominally predicted in Eq. (\ref{power})
because they can vanish 
to the leading order 
due to an even-odd symmetry 
and there can be $\log N$ correction\cite{ALTSHULER,Chubukov}.
However, all planar diagrams obey Eq. (\ref{power}) without $\log N$ correction.

\section{Stability and Anomalous dimension}

The coupling $e$ in Eq. (\ref{a3}) receives quantum corrections only
from the boson self energy due to the Ward identity.
The absence of non-vanishing planar diagrams for boson self energy
beyond the one-loop level 
implies that the one-loop beta function 
is exact in the large $N$ limit\cite{inst}.
Since the one-loop boson self energy has no divergence,
the beta function is zero and the theory is stable
in the large $N$ limit.
The fact that the one-loop result is exact 
is rather remarkable given that the theory
remains strongly coupled even in the large $N$ limit.
This is consistent with an earlier two-loop calculation\cite{KIM94}.
The absence of higher order corrections is 
reminiscent of supersymmetric theories
where certain properties are protected from higher-loop
corrections due to the non-renormalization theorem\cite{SEIBERG}.

In contrast to the boson, 
it is difficult to extract 
detailed dynamical properties 
of low energy fermions on the Fermi surface
even in the large $N$ limit 
because there are infinitely many planar diagrams to be considered.
However, one can attempt to study 
the dynamics of fermion on a general ground.
Here we will show that there is no UV divergence 
in all planar diagrams individually, 
if one uses the one-loop propagators
for the computation of higher-order planar diagrams.
Of course, the one-loop propagator is not reliable
for fermion on the Fermi surface even in the large $N$ limit.
The present approach amounts to summing the one-loop self energy first
and then include the rest of the planar diagrams to examine
whether there is UV divergence or not.

According to Eq. (\ref{scale3}), the scaling dimension of $k$ is given by
\bqa
[k_0] = 1, ~~ [k_x]= \frac{2}{3}, ~~ [k_{y}] = \frac{1}{3}.
\eqa
Every loop contributes scale $2$
and every propagator has scale $-2/3$.
Therefore, the superficial degree of divergence of 
a $E$-point vertex function is given by
\bqa
D_s & = & 2 L - \frac{2}{3} I 
 =  2 \left( 1 - \frac{E}{3} \right),
\eqa
where $E$ is the number of external lines,
$L$ is the number of loops, 
and $I$ is the number of internal propagators.
Here we have used the relations,
$3V  =  E + 2 I$ and  $L  =  I - V + 1$.

There are three kinds of diagrams which are primitively divergent, that is, 
diagrams which have generic divergence at UV 
without divergent sub-diagrams.
The primitively divergent diagrams are 
the 2-point vertex functions (self energies) with $D_s=2/3$ 
and the 3-point vertex function ($D_s = 0$) 
which may have power-law and logarithmic divergences, respectively.

Although the superficial degree of divergence suggests that these diagrams
are potentially UV divergent, there is actually no divergence
for planar diagrams which are dominant in the large $N$ limit.
This can be seen from an argument similar to the one that we used
to prove that all planar boson self energy diagrams vanish.
For planar diagrams, one can assign an internal momentum $k_i$ 
for each single line loop, as we did in Fig. \ref{fig:boson_planar_double_double_momenta}.
(In contrast to Fig. \ref{fig:boson_planar_double_double_momenta},
here we use $k_i$ for all internal momenta to keep the notation for
the following discussion simpler.)
If external frequency is zero,
the integration over $\delta_{k_{i}}$ does not vanish only when
\bqa
|k_{i0}| & <  & \mbox{max}_{j \in i} ( |k_{j0}| ), 
\label{eq:constraint}
\eqa
where $k_j$'s are momenta of fermion propagators
which are parts of the $i$-th loop.
For example, 
for the integration of $\delta_{k_{2}}$ 
in Fig. \ref{fig:boson_planar_double_double_momenta}
to survive,
one should have 
\bqa
|k_{20}| & < & \mbox{max}( |l_{10}|, |l_{20}|, |l_{50}|, |l_{60}|),
\eqa
if $q_0=0$.
These set of constraints can not be simultaneously satisfied for all $i$'s.
If one of the internal frequencies, say $k_{m0}$, is the largest, 
then the $\delta_{k_{m}}$ integration vanishes
because all poles are on the same side in the complex plane.
This implies that the volume of frequency integral vanishes
when the external frequency vanishes.
Since at least one frequency integral has a UV cut-off 
at an external frequency, 
the remaining integrations have 
a reduced degree of divergence 
which is at most $2/3-1 = -1/3$.
Therefore there is no UV divergence in all planar diagrams.

If individually finite planar diagrams can be summed to give a finite result,
the theory is UV finite in the large $N$ limit.
This would imply that the scaling dimension of the fermion field 
given by Eq. (\ref{scale3})
will not be modified by higher-loop diagrams in the large $N$ limit.
However, we emphasize that it is not clear 
whether the summation over planar diagrams 
is convergent or not.
We believe that the true nature of this theory
has not been fully understood.

\begin{figure}[h]
        \includegraphics[height=2.5cm,width=2.5cm]{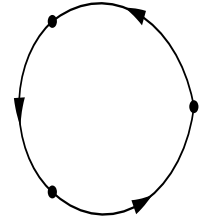} 
\caption{The one-loop diagram for the three-point density correlation function.
The dots represent the density operators.}
\label{fig:D3}
\end{figure}

\begin{figure}
        \includegraphics[height=4cm,width=5cm]{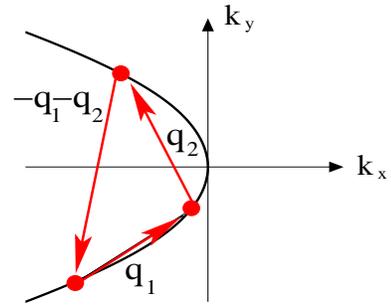} 
\caption{ 
A set of external momenta for which
the three-point density correlation function is enhanced.
If external momenta are chosen so that all of them connect
two points on the Fermi surface, 
it is possible that all internal fermions stay on the Fermi surface.
This gives rise to the divergence $1/\eta$ for the diagram in Fig. \ref{fig:D3} 
if the bare fermion propagator is used.
This can be understood following the same argument given in Sec. III.
There are three internal propagators which can diverge when $\eta=0$.
Since there are only two integrations for the spatial components of
internal momenta, 
a linear divergence survives.
Once the fermion self energy is included, 
the linear divergence is traded with an enhancement factor $N$. 
}
\label{fig:FS_D3}
\end{figure}

\begin{figure}[h]
        \includegraphics[height=6cm,width=4cm]{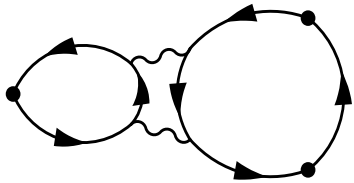} 
\caption{
A three-loop planar diagram which contributes to the leading order three-point density correlation function
when external momenta are chosen so that all of them connect two points on the Fermi surface.
}
\label{fig:D3_2}
\end{figure}

\begin{figure}
        \includegraphics[height=2.5cm,width=4.5cm]{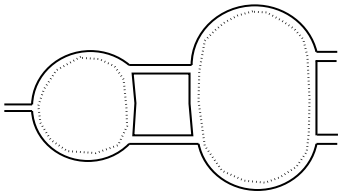} 
\caption{ 
If the external momenta in Fig. \ref{fig:D3_2} satisfy Eq. (\ref{eq:ex}),
every external momentum can be decomposed into two momenta on the Fermi surface.
This allows one to draw the diagram
in the double line representation as is shown in this figure.
There is one loop of solid single line in the double line representation,
which implies that there is an one-dimensional singular manifold 
in the space of internal momenta 
within which all internal fermions stay on the Fermi surface.
As a result, the order of the diagram is given by 
$Q = -4/2 + 2 + [ 1+ 3 -2] = 2$, where we use
$V=4$, $L_f=2$, $n=1$, $E_f=0$ and $E_b=3$ in Eq. (\ref{power}).
}
\label{fig:D3_double_double}
\end{figure}

\section{Correlation function of gauge invariant operators}

Since the fermion operator is not gauge invariant,
the fermion Green's function can not be directly measured.
Therefore it is of interest to study 
how correlation functions of gauge invariant operators
are affected by the fact underlying fermions remain 
strongly coupled in the low energy limit.
Here we consider correlation functions of the density operator, 
$\rho(x) = \psi^*_j(x) \psi_j(x)$.
The two-point correlation function $< \rho(q) \rho(-q) >$ 
is proportional to the self energy of the gauge boson.
As is shown in Sec. III, all planar boson self energy diagrams vanish
except for the one-loop diagram.
Therefore the density-density correlation function shows the 
usual Fermi liquid-like behavior in the large $N$ limit\cite{KIM94}.
Higher order terms become important
only for the $n$-point density correlation function with $n \geq 3$.
Contrary to the two-point correlation function,
higher order diagrams for the $n$-point function with $n \geq 3$ 
do not automatically vanish. 
Since there are more than one external frequency,
the argument used in Sec. III 
to show that all planar boson self energy diagrams with more than one loop vanish
does not apply. 
For example, let us consider the three-point density correlation function given by
\bqa
D_3(q_1, q_2 ) & = & \left< \rho(q_1) \rho( q_2) \rho( -q_1-q_2 ) \right>.
\eqa
For generic external momenta, the leading contribution is given by 
the one-loop diagram shown in Fig. \ref{fig:D3}, 
which is of the order of $N$.
However, for a set of external momenta which connect points on the Fermi surface,
the magnitude of the diagram is enhanced.
If the spatial components of the external momenta can be written as
\bqa
{\bf q}_1 & = & {\bf k}_2 - {\bf k}_1, \nn
{\bf q}_2 & = & {\bf k}_1 - {\bf k}_3,
\label{eq:ex}
\eqa
where ${\bf k}_i$'s are momenta residing on the Fermi surface,
the diagram is enhanced to the order of $N^2$.
For these special external momenta,
there is a channel for all virtual particle-hole excitations
to remain on the 
Fermi surface, 
and there is an enhancement factor $N$.
This is illustrated in Fig. \ref{fig:FS_D3}.
For these external momenta, 
there are infinitely many planar diagrams
which are of the same order.
For example, the diagram in Fig. \ref{fig:D3_2} is of the order of $N^0$
for generic external momenta, but it is enhanced to the order of
$N^2$ if external momenta satisfy Eq. (\ref{eq:ex}), 
as explained in Fig. \ref{fig:D3_double_double}.

\section{Discussions}

\subsection{Comparison with the SU(N) gauge theory}

The genus expansion of the present theory 
is similar to that of the 3+1D SU(N) gauge theory 
in the large $N$ limit\cite{tHOOFT}.
In the SU(N) gauge theory,
all Feynman diagrams 
can be naturally drawn in the double line representation
because the gauge field is in the adjoint representation.
Although the physical origin 
is very different from the present case,
non-planar diagrams with genus $g$ are suppressed by 
the factor $1/N^{2g}$.
One key difference from the present theory is
that there is another dimensionless parameter 
called 't Hooft coupling $\lambda = N g_{YM}^2$
in the SU(N) gauge theory, where $g_{YM}$ 
is the gauge coupling.
This allows for a double expansion of the theory 
in $1/N$ and $\lambda$.
For $\lambda << 1$, the theory is in the perturbative regime.
In the large $N$ limit with a large but fixed 't Hooft coupling $\lambda >> 1$,
planar diagrams with a large number of loops are dominant
and the usual perturbative approach breaks down even in the large $N$ limit.
It has been suggested that 
a more weakly coupled effective description of the theory 
should be a string theory in an one higher dimensional (4+1D) curved space\cite{POLYAKOV1998}.
The AdS/CFT correspondence\cite{MALDACENA,GUBSER,WITTEN} 
is a concrete conjecture of this kind for 
a supersymmetric SU(N) gauge theory.
On the other hand, the present theory with Fermi surface
has no 't Hooft coupling that one can tune in addition to $N$.
To put it otherwise, the effective 't Hooft coupling has been 
set to be of the order of $1$.
This is because there is no dimensionless parameter in the theory
other than $N$ as discussed earlier.
With $\lambda \sim 1$, the theory is still strongly interacting, 
but it is most likely not in the regime where one can use 
a dual gravity description in a weakly curved space-time\cite{Lee09,Hong,Zaanen}.
It would be of great interest to find 
a more weakly coupled description, 
which is likely to be a gauge invariant description, 
for this non-Fermi liquid state.

\begin{figure}[h]
        \includegraphics[height=4.5cm,width=5cm]{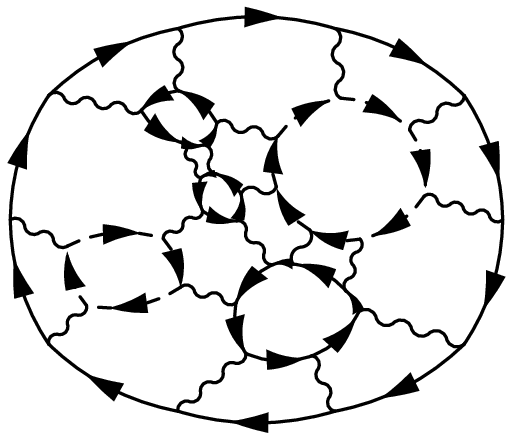} 
\caption{
A planar diagram which include loops of fermions on both sides of the Fermi surface.
The solid (dashed) lines represent the propagator of fermions on 
one (the opposite) side of the Fermi surface.
In order for fermions to remain on the Fermi surface,
two fermions propagators that face each other
should run in the opposite (same) direction
if the two fermions are on the same (opposite) side
of the Fermi surface.
}
\label{fig:vacuum_planar_2}
\end{figure}

\subsection{Extension to multiple patches and applicability of the theory with one patch}

In this paper, we have focused on 
low energy fermion on one side of Fermi surface (one patch).
If Fermi surface is closed, one should consider multiple patches
which include the opposite side  of the Fermi surface
because fermions whose Fermi velocities 
are parallel or anti-parallel
with each other are all strongly coupled 
with the boson in the same momentum region.
The theory which includes fermions with opposite Fermi velocities 
is given by
\bqa
{\cal L} & = & \sum_j \sum_{s=\pm} 
\psi^*_{js} 
( \eta \partial_\tau - i s v_{sx} \partial_x - v_{sy} \partial_y^2 ) \psi_{js} \nn
&& + \frac{e}{\sqrt{N}} \sum_{j,s}  s ~ a ~ \psi^*_{js} \psi_{js} 
 + a ( - \partial_y^2 ) a,
\label{a4}
\eqa
where $\psi_{j-}$ ($\psi_{j+}$) is the fermion 
whose velocity is parallel (anti-parallel) to the $x$-direction.
It turns out that this theory 
is more complicated.
For example, vacuum diagrams which include both $\psi_+$ and $\psi_-$
also contribute to planar diagrams.
An example is shown in Fig. \ref{fig:vacuum_planar_2}.
Those diagrams have the same enhancement factor as those
which involve fermionic loops only on one side of the Fermi surface,
if the curvatures of the Fermi surfaces are 
the same, that is, $v_{+x}/v_{+y} = v_{-x}/v_{-y}$.
A complication arises because 
there is no constraint on internal frequencies 
like Eq. (\ref{eq:constraint})
in the presence of fermions with opposite velocities.
As a result, 
planar boson self energy diagrams do not vanish in general
and there exist UV divergences which are absent in the one patch theory
at least in the large $N$ limit.
One possible scenarios is that 
although there are UV divergences in individual diagrams, 
they cancel with each other and the coupling does not run.
This scenario is consistent with 
the explicit two-loop calculation\cite{KIM94}.
If this is the case, we will obtain a similar picture as the one patch theory.
The question on how to sum all planar diagrams still remains.

If the curvatures on the opposite sides of the Fermi surface 
do not match ($v_{+x}/v_{+y} \neq v_{-x}/v_{-y}$), 
diagrams which has mixed fermion loops 
like the one in Fig. \ref{fig:vacuum_planar_2} has smaller 
enhancement factor because all fermions can not stay on the Fermi surface
because of the curvature mismatch.
In this case one side of the Fermi surface 
which has a smaller curvature will be dominant
and one can focus on one patch as we did in this paper.

\section{Acknowledgment}
The author thanks
Andrey Chubukov,
Piers Coleman,
Matthew Fisher,
Yong Baek Kim,
Hong Liu
and
Olexei Motrunich
for useful discussions.
This work has been supported by NSERC.


\begin{thebibliography}{99}
\bibitem{FL} One can ignore four fermion interactions which induce `non-local' interactions in the momentum space because they are irrelevant in the large $N$ limit.
\end{thebibliography}
\end{document}